\documentclass[preprint,aps,prb,floatfix]{revtex4}
\usepackage{lscape}
\usepackage{epsf}
\usepackage[dvips]{graphicx}
\usepackage{rotating}
\usepackage{bm}
\usepackage{color}
\usepackage{ulem}
\begin{document}

\title{Electronic structure effects in stability and quantum conductance in 2D gold nanowires }

\author{Vikas Kashid,$^{1}$ Vaishali Shah$^{2}$ and  H. G. Salunke$^{3}$} 

\affiliation{
$^{1}$Department of Physics, University of Pune, Pune 411 007, India \\
$^{2}$Interdisciplinary School of Scientific Computing, University of Pune, Pune 411 007, India \\
$^{3}$Technical Physics Division, Bhabha Atomic Research Center, Mumbai 400 085, India
}

\begin{abstract} 

We have investigated the stability and conductivity of unsupported, two dimensional infinite gold nanowires using \textit{ab-initio} density functional theory (DFT). Two dimensional ribbon like nanowires, with 1-5 rows of gold atoms in the non-periodic
direction and with different possible structures have been considered.  The nanowires with $>2$ rows of atoms
exhibit dimerization, similar to finite wires, along the non-periodic direction. Our results show that in these zero thickness nanowires, the parallelogram motif is the most stable. A comparison between parallelogram and rectangular shaped nanowires of increasing width indicates that zero thickness (111) oriented wires have a higher stability over (100). 
A detailed analysis of the electronic structure, reveals that the (111) oriented structures show increased 
delocalization of $s$ and $p$ electrons in addition to a stronger delocalization of the $d$ electrons and hence are 
the most stable.
The density of states show that the nanowires are metallic and conducting except for the double zigzag structure, which, is semiconducting. 
Conductance calculations show transmission for a wide range of energies in all the stable nanowires with more 
than two rows of atoms.  The conductance channels are not
purely $s$ and have strong contributions from the $d$ levels and weak contributions from the $p$ levels.\\

\vskip 0.5in
\noindent
$^2$ Corresponding author. e-mail: vaishali@unipune.ac.in

\end{abstract}

\maketitle

\section{Introduction} 
Structural stability of gold nanowires (NWs) and the resulting effect on conductance is of interest not only from its applicatory value but also for its basic physics studies. Finite size effects, resulting from the  thinning of wires, on the structural stability and electronic properties is an interesting topic for study. Yanson \textit{et al.}\cite{Yanson} and Ohnishi \textit{et al.}\cite{Ohnishi} showed experimentally that gold NWs fabricated between the two electrodes during the thinning process transform successively from quadruple to linear strands. The conductance quanta in these NWs changed from 4 to 1 G$_{o}$ (where, G$_{o}$ = 2e$^{2}$/h). They also found that the (111) and (100) atomic sheets had a stable structure and deformation of the atomic sheets lead to an increase in the conductance. Theoretically, a first systematic investigation in NWs of gold was performed by Sanchez-Portal \textit{et al.},\cite{portal, sanchezd} where-in, the authors observed that zigzag NWs were more stable than single atomic chains (SAC) and other nano-structures. It is well known from Peirel's argument,\cite{peierls} that SAC would be less stable than either bi-atomic chains (BAC) or zigzag structures. Hence, zigzag structures contributed a important finding in Sanchez-Portal's paper. More recently, F. Tavazza \textit{et al.}\cite{tavazza1, tavazza2} have performed a relatively more detailed theoretical investigation of gold wires under tensile stress and strain. They observed that gold wires elongate to form single atom chains (SAC) just before fracture via a series of intermediate structural transformations. The authors investigated seven structures as a starting point and investigated evolution of these into single atom chains under tensile strain. The authors also observed 2D structures are energetically more favorable than 3D structures. However, this study was limited only to the stability of structures under stress and strain. There are no reports in literature on detailed investigations into the electronic origins of quantum conductance of wires different than linear.

Most of the theoretical and experimental investigations published so far are limited largely to structural stability and the measurement of conductance in gold NWs pulled from atomic tips.\cite{dreher, rego, vrodri, rodrigues2, enomoto, ke, bettini} Furthermore most of the investigations are limited to single or double rows of gold NWs. The short length (4-5 atoms) and double row chains have limited potential for practical applications in comparison with somewhat longer atomic chains.
 A further drawback of investigations performed on  NWs suspended in tips is that tips may not be implementable in functional devices. Structural stability of longer NWs under device operating conditions and with respect to the likely atomic processes happening at these temperatures, however will play an important role in the integration of NWs in various technologies. 
 
In the present paper, we intend to investigate stability of infinitely long gold NWs and concomitant  quantum conductance in the wires from a detailed theoretical perspective.  
We have investigated the stability of two dimensional (2D) ribbon like gold NWs of different shapes with increasing number of rows. Besides the geometrical structure, the electronic structure has been analyzed to understand the correlation between the structure, stability and conductivity in these wires. Our results 
show that in the nanoregime the electronic contributions have an important role in determining the stability and conductivity of the NWs. We discuss the increased stability of (111) oriented NWs 
over (100) NWs based on the contributions of $p$ electrons. All 
previous investigations on conductivity in NWs are limited to measuring the conductance, however, the electronic origins of conductance have not been investigated in detail.

We believe, this study will aid further systematic investigations into the effect of different alloying elements on the stability, conductivity and other physical properties of these and similar NWs, an area yet unexplored extensively.
Such investigations especially related with NWs composed of magnetic and non-magnetic elements may reveal novel magnetic 
and conducting properties.
The organization of the paper is as follows. In the next section, we describe the technical details of our simulations on the gold NWs. In Sec. III, we describe our results on the geometric structure and discuss the stability of the NWs based on their electronic structure. The correlation between electronic structure and conductivity is discussed 
in Sec. IV followed by a brief summary of the main results in Sec. V. 

\section{Computational Method}

We have performed total energy calculations based on \textit{ab intio} density functional theory (DFT) 
as implemented in the  Vienna \textit{ab initio} Simulation Package(VASP).\cite{vasp1, vasp2, vasp3} The  Projector Augmented Wave method (PAW)\cite{bloch} and Generalized Gradient Approximation (GGA)\cite{perdew} has been used to describe electron-ion interactions and exchange correlation interactions.   
The NWs are modeled in a periodically repeated unit cell, whose dimensions are chosen 
in such a way that the wires are infinite and periodic in the $x$ direction with 12 \AA ~ vacuum spacing in 
$y$ and $z$ directions. A plane-wave energy cut-off of  $229.9 $ eV ($13.2$ Ry) was used. 

The energy convergence of linear wire was tested with respect to the number of k-points used for integrating the Brillouin Zone. 
A $13 \times 1 \times 1$ mesh of k-points in the Monkhorst Pack scheme gave reasonable convergence in the total
energy. The same mesh of k-points has been used for all other structures. Atomic relaxations are performed by conserving the shape and volume of the unit cell. The NW structures have been optimized using the Residual Minimization (RMM-DIIS) \cite{puley} method and the Conjugate gradient (CG) \cite{teter, stich} method. The structural optimizations have been performed until the total forces on individual atoms are less than $0.01$ eV/\AA.~ 
The binding energy per atom (E$_{B}$) for gold NWs is calculated as \[ E_{B} = \frac{[ E_{T} - n E_{atom}]}{n} \] \\
where, E$_{T}$ is the total energy of the NW, E$_{atom}$ is the energy of single gold atom and $n$ is the total number of atoms.

In order to understand the electronic and conduction properties of the NWs, we have performed band structure calculations. 
For the linear wire,  the Brillouin zone integration has been performed from $\Gamma$(0,0,0) to X(0.5,0,0) direction with 10 equidistant kpoints. 
Except for the linear wire, all other geometries are two dimensional.  Considering this, the two dimensional 
Brillouin zone has been integrated along $\Gamma$(0,0,0) to X(0.5,0.5,0)- direction using 10 equidistant kpoints. 
In order to understand the contributions of the electrons in the {\it s}, {\it p} and {\it d} states to the energy levels in the 
band structure, the `fat band technique' was applied. The technique has been developed and implemented by Jepsen {\it et al.} \cite{fat} in the LMTO-ASA tight binding method. We have developed a similar technique in VASP by giving
a width to the electronic energy band that is proportional to the sum of the corresponding orthonormal orbital(s).
The details of the technique of the `fat band method' with VASP package are described in V. Kashid {\it et al.} \cite{kashid}
We have also performed spin polarized DFT calculations for the infinite NW geometries. These calculations show that infinite NWs are non-magnetic in nature.

The conducting properties of gold NWs are investigated based on
the Landuaer-B\"{u}ttiker formalism, \cite{landauer} using the method of plane waves, \cite{oetzel} For each nanowire, the conductance is calculated as
 
\[G(E) = \frac{2e^{2}}{h} T(E)\]

where, T(E) is the transmission function at a particular energy and calculated as  \[T(E)=  N(E),\] 
Here, $N(E)$ are number of states. The number of states can be directly obtained from the band structure of the gold 
NWs.

\section{Results and Discussion}

\subsection{Structure}

Structural optimizations were performed on fourteen different structures of gold nanowires with 1-5 rows of atoms. 
Except for the linear (LR) nanowire, all other NWs studied are two dimensional, with increasing number of rows of Au 
atoms, in one of the non-periodic directions. 
Amongst 14 unconstrained NW structures we investigated, some structures have been previously reported in the literature. The linear NW has been studied in great detail as free standing\cite{Jun,sanchezd,portal,ke,Torres} as well as suspended between tips.\cite{dreher,kurui,tavazza1,tavazza2,vrodri,Okama} Unconstrained structures like zigzag (ZZ), rectangular (RT-1), parallelogram (PR-1) and double zigzag (DZZ) structures have also been investigated by other researchers.\cite{Jun,portal,sanchezd,kurui,fioravante} More recently, Tavazza \textit{et al.} have found that 2D structures like ZZ, PR-1, PR-2, DZZ and centered hexagonal (CHX) are formed intermediately during the thinning process.\cite{tavazza2}

We optimized all the NW geometries by varying the bond lengths between the gold atoms in the periodic direction. 
Figure 1 shows the structures of NWs investigated. The structures in Fig. 1 are ordered according to the increasing 
number of rows of atoms in the 2D NWs and for a particular number of rows, the structures are shown with increasing stability. 
In Fig. 1, for a given 2D row of atoms, we have nearly exhausted all structural possibilities. For example, in Fig. 1(d), with four rows of atoms, four structural possibilities are applicable. Thus, hexagonal (HX), DZZ, triple rectangle (RT-3) and triple parallelogram (PR-3) structures are investigated. 
In most of the wires, the optimized structures have been found to have primarily, zigzag, rectangular and parallelogram motifs or their combinations.

\subsubsection{Single and Double row NWs}
The single row linear NW shown in Fig. 1(a) is well studied. The possible simplest 2D NW with 2 rows of atoms are shown in Fig. 1(b) viz., zigzag (ZZ), rectangular (RT-1)  and parallelogram (PR-1) (referred as `zigzag under compression' by Portal \textit{et al.} \cite{portal, sanchezd}). We have investigated these wires for verification and completeness of our calculations.  Most of the literature on linear NW is for the wires suspended between the tips. Linear NW has been found to be highly stable under stretching forces and
has a quantum conductance of 1G$_o$ at room temperature.\cite{skoro,Jun,Torres,Ohnishi, Yanson}
Our optimized linear wire with an uniform Au-Au bond length of 2.61 \AA, is in good agreement with the bond length of 
2.60 \AA~ reported  by Zarechnaya \textit{et al.}\cite{zarechnaya} 
Unlike finite gold NWs, \cite{xiao, presenjit} the infinite linear NWs do not show variation in the bond length or dimerization, in agreement with, the calculations by Xiao {\it et al.} \cite{xiao} and Nakamura {\it et al.}\cite{Jun}

Our optimized ZZ structure has a bond length of 2.55 \AA ~  and bond angle of $131.5^{\circ}$.  This agrees very well with the bond angle of $131^{\circ}$ calculated by D. Portal \textit{et al.} \cite{portal} Our calculated distance between Au$_{1}$ and Au$_{3}$ atoms is 4.66 \AA, which, compares quite well with 4.62 \AA~ obtained by D. Portal \textit{et al.} \cite{portal} 
that demonstrate the binding energy plot has two minima as a function of bond angle for the ZZ structure.
Our calculations are in confirmation with earlier observations that the ZZ structure is a metastable structure when compared with PR-1 structure. These results were arrived from the binding energy plot as a function of bond angle. \cite{portal,sanchezd}
 The ZZ structure is at the higher minimum (-1.90 eV) and the PR-1 structure is at the lower minimum (-2.22 eV) of the binding energy plot.   
The RT-1 structure  has a length of 2.65 \AA, width of 2.68 \AA~ and the bond angles of 90$^{\circ}$. The relativistic calculations by Nakamura \textit{et al.} \cite{Jun} have obtained 2.59 \AA~ as the length and width of the rectangular wire. \cite{Jun}
The PR-1 structure, is a more compact structure than the zigzag structure with all acute angles between 
the gold atoms. The length and width of parallelogram is 2.67 \AA~ and 2.76 \AA~ respectively. An average bond length of 2.76 \AA~ for parallelogram shaped NW has been reported earlier by Tavazza \textit{et al.} under elastic deformation in between the tips.\cite{tavazza2} The bond angle analysis shows that the parallelogram is formed of two isosceles triangles in disagreement with the structure reported by Fioravante \textit{et al.}, \cite{fioravante} which is 
made up of equilateral triangles. The preference for isosceles triangles over equilateral triangles is on account of symmetry breaking along the non-periodic $y$ direction. The PR-1 structure is found to be the most stable double row NW, in agreement, with the earlier calculations by Portal \textit{et al.}\cite{portal, sanchezd} and Nakamura \textit{et al.}\cite{Jun} 

\subsubsection{Triple row NWs}
We have examined three, triple row NW structures, double rectangular (RT-2), pentagonal (PNT), and double parallelogram (PR-2) as shown in Fig. 1(c). 
We have also performed optimization of another three row geometry, i.e., the rhombus structure (not shown in the figure).
We find that further lattice optimization of the infinite length rhombus structure leads to the energetically lower 
PR-2 structure. Unlike the ZZ structure, which has a distinct minima in the binding energy
versus lattice separation diagram, the rhombus structure does not reside in a distinct potential well. 
The optimization path smoothly curves down to the PR-2 structure. Recently, a single rhombus structure held between parallelogram NWs on both sides was observed as an intermediate structure during compression of linear NW.\cite{tavazza2} However, we would like to note that, so far, there have been no experimental reports of two dimensional rhombus shaped NW.\cite{kurui}

The RT-2 structure, is made up of uniform rectangles stacked one on top of another, with a 
length of 2.70 \AA, and width of 2.62 \AA. 
The initial geometry set up for the PNT NW was made up of regular pentagons. Geometry optimization 
of this regular structure lead to a structure that has an admix of a RT-1 NW and a 
PR-1 NW as shown in Fig. 1c (ii). The horizontal bond length is enhanced by 0.05 \AA~ and 0.03 \AA~ in comparison with their equivalent bond lengths in RT-1 and PR-1 structures respectively. The diagonal bond length in PNT is reduced by 0.02 \AA~ in comparison with PR-1 structure. The bond angles between Au$_{2}$-Au$_{5}$-Au$_{4}$ and Au$_{2}$-Au$_{4}$-Au$_{5}$ are $60.5^{\circ}$ and $59.0^{\circ}$ respectively. The corresponding angles in the PR-1 structure are $61.2^{\circ}$ and $57.6^{\circ}$, respectively.

The PR-2 structure is the most stable three row geometry. In the PR-2 NW, Au$_{1}$-Au$_{2}$ have a bond length of $2.74$ \AA~ and Au$_{2}$-Au$_{5}$ have a bond length  $2.80$ \AA.~ The bond angles between Au$_{1}$-Au$_{2}$-Au$_{6}$ and Au$_{2}$-Au$_{6}$-Au$_{1}$ are $61.4^{\circ}$ and $60.3^{\circ}$ respectively. Our analysis of bond angles shows that PR-2 structure is comprised of slightly distorted equivalent triangles.  

\subsubsection{Quadruple row NWs}
Fig. 1(d) shows the four row NW structures considered, namely, hexagonal (HX), double zigzag (DZZ), triple rectangular (RT-3), and triple parallelogram (PR-3).
The optimized HX structure has bond lengths between  Au$_{1}$-Au$_{2}$, Au$_{2}$-Au$_{3}$ and Au$_{2}$-Au$_{15}$ as $2.58$ \AA, ~$2.58$ \AA, ~and $2.69$ \AA~ respectively. The structure is a slightly distorted hexagon with bond angles of 115.1 and 129.8$^{\circ}$.
The DZZ structure is made up of two zigzag NWs, with one placed on top of another one.  The 
bond lengths between the atoms forming the zigzag shape, Au$_{1}$-Au$_{4}$, Au$_{2}$-Au$_{3}$, Au$_{4}$-Au$_{5}$, Au$_{3}$-Au$_{6}$  are $2.67$ \AA. The bond lengths between the atoms Au$_{2}$-Au$_{4}$, Au$_{4}$-Au$_{6}$ are 2.65 \AA. Bond angle between Au$_{1}$-Au$_{4}$-Au$_{2}$ is 65.4$^{\circ}$, whereas, the bond angle between the atoms Au$_{1}$-Au$_{4}$-Au$_{5}$ is 113.9$^{\circ}$. In the DZZ structure, the bond lengths have stretched by 0.12 \AA ~in comparison with the ZZ structure and the bond angles have reduced.

The RT-3 wire is interesting, as, it is the first NW geometry that exhibits 
dimerization in the $y$ direction. Although dimerization in finite linear wires is well observed and reported,\cite{xiao} dimerization in thicker wires has not been reported so far. The length of the rectangles in RT-3 NW is 2.68 \AA ~. The widths of the 
top, middle and bottom rectangles are 2.65 \AA, 2.61 \AA ~and 2.65 \AA, ~respectively. The dimerization has been explained on the basis of non-uniform
potential experienced by the atoms in the finite length direction. No dimerization is observed in
the infinite length direction in gold NWs on account of every atom experiencing a uniform 
potential.\cite{Jun,xiao}
The PR-3 structure is formed of parallelograms that have a uniform length of 
2.72 \AA. The widths of the two terminal parallelograms
are 2.75 \AA, which have expanded by 0.01 \AA ~in comparison with the PR-2 NW. The width
of the central parallelograms is 2.74 \AA, same, as that of the PR-2 NW. 
\subsubsection{Quintuple row NWs}
We have investigated three, five row NW structures, centered hexagonal (CHX), quadruple rectangular (RT-4) 
and quadruple parallelogram (PR-4) as shown in Fig.1(e). 
The CHX wire is generated from the HX wire (shown in Fig. 1d (i)) by placing a gold atom 
at the center of each hexagon. The additional atom in the center, expands the length of each side of hexagon 
by 0.12 \AA. The bond angles are observed to expand in one case from 115.1 to 118.3$^{\circ}$, while the other contracts 
from 129.8 to 123.4$^{\circ}$ in comparison with that of the HX NWs.

The RT-4 structure shows dimerization in the non-periodic direction like the RT-3 NW. The terminal
rectangles have a width of 2.66 \AA ~and the two central rectangles have a width of 2.63 \AA. The 
terminal and central rectangles have a uniform length of 2.66 \AA.  
The PR-4 structure is made up of slightly distorted equilateral triangles. The length of the 
parallelograms is 2.73 \AA. The width of the terminal parallelograms is 2.76 \AA ~whereas, that of the two central 
parallelograms is 2.74 \AA. 
Similar to the RT-4 NW, dimerization in the length along the non-periodic direction
is observed.

In general, as the number of rows in the NW increases, the bond lengths between the gold atoms along the 
length of the NW are seen to increase, whereas, the bond lengths perpendicular to the length decrease. 
Linear NW has the smallest bond length (2.61 \AA) and the  PR-4 has the largest bond lengths (2.73 \AA).  
This bond length is smaller by 0.22 \AA ~in comparison with the bulk bond length of 2.95 \AA.
The bond angles show an opposite trend. The bond angles are decreasing with an increasing number of rows in the NWs. In the PR shaped NW's, the bond angles are converging with increasing number of rows to their bulk value of 60$^{\circ}$.
Within a particular number of rows, the bond lengths of the less stable NW structures are smaller than
the bond lengths of the most stable NWs.

Similar to finite gold NWs,\cite{xiao}  dimerization along the finite second dimension of the wire,  is observed in RT-3, PR-3, RT-4 and PR-4 NWs. 
We would like to note that the structure of RT-3 and RT-4 wires is similar to  
zero thickness nanoribbon of (100) surface 
and that of PR-3 and PR-4 wires is similar to a zero thickness nanoribbon 
of (111) surface.

\subsection{Stability}

The stability of linear wires has been investigated earlier by 
Nakamura \textit{et al.} \cite{Jun} and Portal \textit{et al.}\cite{portal}  Their calculations have shown that the  PR-1 shaped 
wire is more stable over the ZZ and RT-1 wires. These three  NWs are made up of a double row
of atoms and have a higher stability in comparison with the LR wire. The stability of the ZZ wire over the 
LR wire is argued to be on account of the $d$ levels in the ZZ NWs being situated well below the Fermi level,
in comparison with, that of the LR wire. This raises a few interesting questions related with zero thickness
NWs. Will the stability of the NWs increase as one increases the width i.e. the number of rows in the 2D NWs? Is  
stability related to the location of the $d$ levels with respect to the Fermi level?

In order to investigate this, we have listed the NWs in ascending order of stability in Table I. 
The number of atomic rows in the non-periodic direction are also listed. %The bulk cohesive energy is reported for comparison. 
In agreement with the earlier calculations,\cite{portal} we observe that the LR wire is less stable.
The ZZ NW is stable over linear NW by 0.2 eV  and the PR-1 wire is stable over ZZ and RT-1 wire by 0.32 eV and 0.13 eV, respectively. 

In the three row NW structures, the PR-2 wire is more stable over the RT-2 and PNT wires by 0.15 eV and 0.07 eV, respectively. 
In the four row structures, the DZZ is stable by  0.31 eV in comparison with the HX NWs.
This DZZ NW  has been considered to be a ``magic structure'' adopted by NWs during the thinning 
process.\cite{fioravante} 
Interestingly, this four row structure is less stable not only in comparison with the three row PNT and PR-2 NWs but also in comparison with all other four row (except hexagonal) and five row NWs. 
In the five row structures, the CHX wire is less stable by 0.22 eV in comparison with the PR-4. It is most interesting to note that the RT-4 NW is stable over the CHX wire by 0.01 eV. 

It is clearly evident from the results in Table I, that the stability of the NWs does not increase linearly 
with an increase in the number of rows in the structures. Figure 2 shows gold NWs sorted on the basis of the stability with the increasing number of rows. It can be seen that the HX wire,  with 4 rows, is less stable than the double row NWs. (We have noted earlier that the ZZ is not stable structure, rather the PR-1 structure
is stable) The PR-2 NW, with 3 rows, is highly stable in comparison with most of the 4 row and 5 row NWs. 
What is more interesting is that the double, triple and quadruple parallelogram NWs are the most stable not only 
within a particular number of rows but even across them. The next stable NWs are the rectangular shaped NWs.  
However, as one increases the number of rows, the energy difference between the rectangular and parallelogram 
NWs is seen to gradually increase as 0.13 eV (double row), 0.15 eV (triple row),  0.18 eV (quadruple row) and 0.21 eV (quintuple row). Hence, we conclude that the parallelogram motif is the most stable motif for 2D gold NWs.
We performed additional calculations by increasing the number of rows to 18 for only the parallelogram and rectangular shaped NWs. In addition, we have obtained the binding energies of a single layer surface of gold having (111) and (100) orientations as -2.86 eV/atom and -2.58 eV/atom respectively. Our results in inset of Fig. 2 show that the binding energies of higher row parallelogram and rectangular NWs are converging to their zero thickness film limit. These results indicate that (111) single layer films will have a higher stability over (100) films.

To further investigate the stability of the gold NW structures, we have performed a nearest neighbor analysis of the structures based on the charge density distribution (charge density plots not included here). The spatial distribution of electronic charge is observed to be symmetric along the bonding direction of atoms in the NWs. The charge density distribution of the NWs shows that as one goes from linear to PR-4 structure, there is an increased sharing of electrons between the atoms. The homogeneous distribution of electron density in all the NWs investigated here, is indicative of metallic bonding in the NWs.

The number of nearest neighbors of the different atoms in the NWs are listed in Table II  along with the number of bonds per atom.  In the nearest neighbor column of Table II, more than one entry is listed to account for different
 atoms having different number of nearest neighbors, depending on whether they belong to the edge rows or one of the central rows.  Within a particular number of rows of atoms, the NW with the largest number of bonds per atom 
 is found to be the most stable. In the double row NWs, PR-1 with 2 bonds per atom is most stable.
In the triple row structures, PR-2 with 2.33 bonds/atom is most stable. In quadruple row structures, PR-3 with 2.5 bonds per atom is most stable and in quintuple row structures the PR-4 with
2.80 bonds/atom is the most stable.
An increased number of bonds between the atoms tends to increase the stability of the NWs. However, 
this bonding picture does not explain the higher stability of RT-2, PNT and DZZ with less number of bonds per atom than the parallelogram NW. Thus, it is important to consider other electronic effects when understanding the stability of the NWs.

Figure 3 shows the electronic band structure (BS) in the left panel and the density of states (DOS) in the right panel for three NW structures viz., LR, RT-4 and PR-4 representing the least stable and most stable among rectangular and parallelogram structures respectively.
 The horizontal dotted line represents the Fermi level. In all the 14 NWs, the major contributions to the density of states are from the $d$ electrons. Hence, only the $d$ contributions to the density of states and the total density of states are shown in the figure.
Our analysis of DOS  shows that all wires are metallic except for the DZZ which has a gap of 1.04 eV 
at the Fermi level. This is the only wire that exhibits semiconducting characteristics, in agreement with the calculations by  Fioravante \textit{et al.} \cite{fioravante}
The DOS shows that within a particular number of rows, the most stable NW has a higher delocalization
of $d$ electrons. 

In the investigations on the stability of LR and ZZ wires, it has been argued that in the LR wire the 
$d$ states are spread till the Fermi level destabilizing these wires in comparison with the ZZ wires. \cite{portal}
Nakamura {\it et al.} \cite{Jun} and Portal {\it et al.} \cite{portal} have accounted
the stability of the ZZ wire to the separation between $s$ and $d$ bands near the Fermi level and the
pulling of the $d$ bands below the Fermi level. 
The calculated DOS of 3, 4 and 5 row NWs show that the $s$ and $d$ states are spread until the Fermi 
level and yet the wires have an enhanced stability over the zigzag wire. 
We have also calculated and listed in Table I, the location of the 
peak of $d$ bands from the Fermi level, and the width of the $d$ bands below the Fermi level. The most stable RT-4, PR-2, PR-3 and PR-4 NWs have their $d$ band peaks closer to the 
Fermi level in comparison with the less stable RT-2, PNT and CHX NWs.  These
results clearly show that the location of the $d$ band peak is not an indicator of the stability of the wires. 

The width of $d$ bands, however, is related with the stability of the wires. In general, the width of the $d$ bands is seen to increase with increased stability. An increase in the width of $d$ bands translates to an increased delocalization of the $d$ electrons. Hence, geometries exhibiting an increased delocalization of the $d$ orbital electrons are the most stable NW structures.
As seen from Table I, the width of the  $d$ bands in the PR-1 wire is less than that of the RT-2, PNT and DZZ wire. This explains the stability of the RT-2, PNT and DZZ wires over the parallelogram NW.
Further, in the case of rectangular NWs, the RT-2 with a larger width of $d$ bands is less
stable in comparison with RT-3 and RT-4 NWs, which have a smaller width of $d$ bands.
Our detailed investigations on the orbital contributions in the DOS of the NWs revealed that besides 
the $d$ bands, the $s$, $p_{x}$ and $p_{y}$ bands play a role in the stability of the wires. With an increase in the 
number of rows in the NWs, the $s$ and $p_{x}$  orbitals exhibit increased localization and enhancement in the peak 
heights. The $p_y$ electrons, on the other hand, show increased delocalization with an increase in the number 
of rows. This increased delocalization of the $p_y$ electrons leads to higher stability in triple and quadruple 
rectangular NWs, in spite of having somewhat smaller width of $d$ bands.  Although the $p$ contributions to 
the density of states are one-tenth of that of the $d$ states, they play an important role in the stability of 
2D NWs.

An analysis of all the parallelogram shaped NWs and their comparison with rectangular shaped NWs reveals that 
all electrons show increased delocalization in the parallelogram NWs (not shown in the figure). The peaks of the DOS of $d$ electrons are at 
the same energies in the parallelogram and rectangular NWs. The $p_x$ and $p_y$ electrons are more delocalized in the parallelogram shaped NWs as compared to that of rectangular NWs. The $s$ electrons show enhanced delocalization 
and enhanced peaks  in comparison with rectangular NWs. The contributions from $s$ electrons are significant (one-third of that of the $d$ electrons). The peaks of the $s$, $p_x$ and $p_y$ levels in the DOS are slightly more shifted 
below the Fermi level in comparison with the peaks of these states in the rectangular NWs.
The delocalization of $s$, $p_x$ and $p_y$ electrons in addition to the delocalization of $d$ electrons 
accounts for the increased stability of these wires in comparison with that of rectangular shaped NWs.

\subsection{Conductivity}
All experimental and theoretical research on conductance in gold NWs reports conductance in quanta of G$_{o}$. The electronic contribution to the conductance behavior has not been investigated in detail except for the linear nanowire.\cite{dreher,Mares,ke,rego} In this section, we present correlation between electronic structure and conductance behavior of 2D gold NWs based on `fat bands' analysis. Such analysis, in general will be useful for other kinds of nanowires as well. 
The conductance in 2D nanowires is calculated based on the Landuaer-B\"{u}ttiker (LB) formalism\cite{landauer} as implemented by Oetzel \textit{et al.} for infinite chains.\cite{oetzel} This formalism has been developed for 
infinite one dimensional chains, where the effects of leads are negligible as their resistance is negligible in
comparison with the wire resistance. Our results indicate that, the formalism works for somewhat wider 
chains as well.
We have obtained the transmission coefficients using LB formalism for the 14 gold NW structures.

For linear NW, the transmission at the Fermi level shows a jump from 3G$_o$ to 1G$_o$ as shown in Fig. 5(a), in agreement with, the earlier reports on conductance calculations.\cite{thygensen,Jun,oetzel} It has been shown by Dreher \textit{et al.}, \cite{dreher} that the stretched linear NW of two atoms suspended between tips has single conductance channel mainly from 6$s$ electrons. At the end of the stretching, the linear NW is of 4 atoms and shows 2 or 3 channels for conductance, which are due to participation of $d$ electrons. Our fat bands analysis on unconstrained NWs plotted in Fig. 4(a) shows that the single conduction channel in linear NW is made of $s$ orbital electrons hybridized with $p_{x}$, $d_{xy}$, $d_{xz}$ orbitals. The $s$, $d_{xy}$ and $d_{xz}$ orbitals have significant contributions in conduction with $p_{x}$ playing a minor role.

Most experimental and theoretical work on gold NWs \cite{thygensen,Jun,oetzel,dreher,kurui} demonstrates that the decrease in the conductance closely follows the decrease in the cross section of NWs during their thinning process. As the cross section of the NW decreases, the conductance decreases to 1G$_{o}$. (The cross section is determined as the number of atomic rows multiplied by the area per atomic row. \cite{kurui}) In analogy with this, in 2D structures, thickness of the NW will be related with the number of rows of atoms in one of the non periodic directions. 
Our conductance calculations listed in Table-II demonstrate that the number of conduction channels are not directly related with the number of rows of atoms. Structure of the NW plays a prominent role in the conduction behavior. HX, DZZ and PR-1 with more than one row of atoms have zero conductance at Fermi level. CHX and ZZ structures, which are somewhat more stable than linear NW have same conductance as linear NW at the Fermi level.
 The most stable rectangular and parallelogram NWs viz., RT-4 and PR-4 does not show any block in the transmission for the complete energy range. Another interesting feature observed is that in rectangular NWs, the conductance is related with the number of rows whereas, in parallelogram NWs, the conductance is less than the number of rows of atoms, in agreement with Kurui \textit{et al.}\cite{kurui} We give an explanation for this based on density of states analysis subsequently.

To understand the origin of conductance in different structures, we have performed fat band analysis of the 14 NWs. In Fig.4, we have shown only 3 representative structures viz., the most investigated LR, second most stable RT-4 and highest stable PR-4. 
In general, we observe that in all gold NWs, the conductance channels are formed by hybridization of $s$, $d_{xy}$, $d_{x^{2}-y^{2}}$, $p_{x}$ and $p_{y}$ orbitals. The $s$, $d_{xy}$ and $d_{x^{2}-y^{2}}$ orbitals have major contribution in the conductance with minor $p_{x}$ and $p_{y}$ orbital characters.
All parallelogram shaped NWs have strong $s$ and $d_{xy}$ character except for PR-1 structure, whereas, in rectangular shaped NWs, the $s$ and $d_{x^{2}-y^{2}}$ orbital gives major contribution at Fermi energy.
As the number of rows in parallelogram NWs increases in non periodic direction, the $d_{x^2-y^2}$ orbital character is slightly enhanced.
Similar to parallelogram NWs, $d_{xy}$ orbital character in rectangular shaped NWs, is slightly enhanced as the number of rows of atoms are increased.  
The contribution from $s$ in the conduction channel are reduced from RT-2 to RT-4 and are the same as RT-2 for all other higher row rectangular NWs. All rectangular NWs show almost same contribution of $d_{x^2-y^2}$ orbital character at Fermi energy with the increasing number of rows in the structure.

In order to investigate the dependence of conduction on number of rows in rectangular and parallelogram NWs, we performed detailed analysis of DOS for these NWs. It has been observed that in the rectangular NWs
the $s$, $p_{x}$, $p_{y}$ and $d_{x^2-y^2}$ electrons have more contributions at the Fermi level. In the parallelogram
shaped NWs, the peaks of the $s$, $p_{x}$ and $p_{y}$ orbitals are more shifted below the Fermi level in comparison with
the rectangular shaped NWs. In both the rectangular and parallelogram shaped NWs, the positions of the peaks of the rest of the $d$ states
are identical. In the rectangular shaped NWs, the contributions from $d_{x^2-y^2}$ electrons at the 
Fermi level are greater than that of parallelogram NWs and contributions from $d_{xy}$ electrons are lesser than 
that of parallelogram NWs. The contributions from rest of the $d$ states are zero at the Fermi level in both types 
of NWs. Thus in the parallelogram shaped NWs, the more bound $s$, $p_{x}$ and $p_{y}$ electrons 
become less available for conduction, but at the same time, they lead to an increase in the stability of the NWs. 
The enhanced contributions from  $s$, $p_{x}$, $p_{y}$ and $d_{x^2-y^2}$ electrons leads to an increased conductivity
in rectangular shaped NWs. From the transmission histograms and the band structure of the rectangular and parallelogram shaped wires, it can be inferred that the jumps in conductivity are related with the number of 
the confined $s$, $p$ and $d$ electrons contributing in the transmission at that particular temperature.

\section{Conclusion}

We have performed detailed investigations on 2D ribbon like gold NWs of 1-5 rows of atoms. Amongst structures with same number of rows, the wires with a parallelogram motif are the most stable. Our study reveals that as the number of rows increase, the bond lengths increase and the bond angles decrease towards their bulk values. Gold NWs $> 3$ rows of atoms, exhibit dimerization in the non periodic direction similar to that observed in finite NWs.

Furthermore our study of the stability analysis of the fourteen structures reveals that the linear wire is the least stable and the quadruple parallelogram NW is the most stable. Contrary to conventional logic, the stability of the NWs does not increase merely with an increase in the number of rows in the NWs but the structure plays an important role. Our calculations demonstrate that the four row hexagonal NW has lower stability than double row pentagonal NW. Some other four and five row NWs also have a lower stability than the triple row PR-2 NW. These results demonstrate that, besides the increased coordination with increasing number of rows of atoms, the geometrical structure plays an important role in determining the stability of NWs. Structural shapes that increase the delocalization of $s$, $p$ and $d$ electrons show an enhanced stability. 

We find that the stability of rectangular motif is second to that parallelogram motif and the difference in their energies increases with increasing number of atomic rows in the non-periodic direction. The quadruple rectangular and quadruple parallelogram are similar to zero thickness nanoribbons of (100) and (111) surfaces, respectively. Our results for NWs with extended rows of atoms until the zero thickness limit show that (111) oriented single layer films have higher stability than (100) single layer films.

Our electronic structure investigations into the stability of 2D nanowires imply that if experimental thinning process were to be carried out on thicker gold NWs then it would result in observation of sequence of PR-4 structure followed by PR-3 structure to PR-2 structure and then to LR structure. Unstable single atom linear chains would be observed in the end just before breaking. In fact, our conclusions have been partly verified in available literature. Recent work by Kurui \textit{et al.}\cite{kurui} observed stepwise triple, double and single strands of (111) and (100) atomic sheets during the thinning process. The theoretical investigations by Tavazza \textit{et al.} \cite{tavazza2} found four stable intermediate atomic configurations during the elongation process, namely, parallelogram (mentioned as `zigzag'), centered hexagonal (mentioned as `hexagonal'), triangular (110) and triangular (111).

The Landuaer-B\"{u}ttiker transmission coefficients of the NWs show that all wires are conducting at 0K 
except for the hexagonal, double parallelogram and double zigzag. Our calculations on the 1-5 row structures show that stability and conductance
are inversely related. The less stable NWs show higher available conductance channels in comparison 
with the more stable NWs. The conductance in 
rectangular wires is equal to, whereas, in parallelogram shaped wires it is less than the number of rows of atoms in the wires.
The band fattened electronic structure of all NWs shows that the conducting channels are not
purely $s$, but have contributions from $p_{x}$, $p_{y}$, $d_{xy}$ and $d_{x^2-y^2}$ electrons. The jumps in the conduction channels are correlated with the electronic band structure of the NWs.
All the wires whether conducting or non-conducting exhibit {\it spd} hybridization. We conclude from the calculations of quantum conductance that supported but otherwise unstable gold NW ribbons could be more conducting than their structurally stable counterparts.

This study serves as the basis for investigating from first principles the 
stability and conductivity of different alloyed NWs. The effects of the different alloying 
elements and their concentration in Au NWs is under investigation. Such 2D NWs will be 
very attractive candidates for (bio)-sensing and (bio)-electronic applications and may also be 
suitable for non-invasive transdermal drug release applications.

\section{Acknowledgments}

This work was performed using BARC mainframe supercomputers (ajeya and ameya). Dilip G Kanhere and Stefan Bl\"{u}gel are gratefully acknowledged
for stimulating discussions. V. S. would like to acknowledge 
funding support from Dept. of Science and Technology, Govt. of India.  V.K. acknowledges BARC for financial 
support. 
V. K. and V. S. wish to gratefully acknowledge the kind hospitality of Bioinformatics Center, University of 
Pune and Institute of Bioinformatics and  Biotechnology, University of Pune, during the course of this work.

\bibliography{reference}
\bibliographystyle{apsrev} 

\newpage 
\textbf{Captions of Tables and Figures}\\

$\mbox{FIG. 1}$: The optimized geometries of the  gold nanowires  with increasing number of rows (1-5) of 
atoms. Within any row, the geometries are arranged in ascending order of stability with (i) depicting the least stable and (iii)/(iv) depicting the most stable structures. \\

$\mbox{FIG. 2}$: Gold nanowire structures shown on binding energy plot as a function of number of rows. Figure in the inset shows extended binding energy curve for parallelogram and rectangular nanowires to 18 rows of atoms. Red filled squares show binding energy per atom for parallelogram motifs, whereas empty blue squares show binding energy per atom for rectangular motifs. \\

$\mbox{FIG. 3}$: The electronic band structure and density of states (DOS) for (i)LR, (ii)RT-4 and (iii)PR-3 NWs. In each plot, 
the left panel shows electronic band structure and right panel shows the corresponding DOS with the Fermi level 
set to zero. The dotted line represents the Fermi level. The blue line represents the DOS of the $d$
electrons and the red line represents the total DOS. \\

$\mbox{FIG. 4}$: The electronic band structure close to Fermi level for gold nanowires using the fat band technique. In all figures, Fermi level is set to zero and shown by the dotted line. The red colored bands are the fattened bands of the band structure that show the contribution from that particular orbital electrons. (a)Linear (LR), (b)Quadruple rectangular (RT-4) and (c) Quadruple parallelogram (PR-4).\\

$\mbox{FIG. 4}$: Transmission histograms of (a) linear (LR), (b) Quadruple rectangular (RT-4) and (c) Quadruple Parallelogram (PR-4) \\

$\mbox{Table I}$: The location of $d$ band peak with respect to Fermi level and the 
width of the $d$ bands of the gold nanowires, arranged according to increasing order of stability. \\

$\mbox{Table II}$: The binding energy (eV/atom), number of nearest neighbors, the number of bonds formed per atom in the gold nanowires and number of conduction channels of gold nanowires shown in Fig. 1. The number of rows of atoms in the nanowire structures are also listed for comparison. \\

% ------------------ Table-1 -----------------------
\newpage
\begin{center}
\begin{table}[h!b!p!]
\begin{tabular} {l  c  c  c } \hline \hline
Geometry & Number \hskip 0.2in & \hskip 0.1in $d$ peak & \hskip 0.2in width of \\ 
     & of rows & \hskip 0.1in  with respect & \hskip 0.2in $d$ bands \\ 
   &  & \hskip 0.1in to E$_{f}$ &  \\ 
\hline
\hline
Linear (LR) & $1$ & $-1.12$ & 5.58 \\

Zigzag (ZZ) & $2$ & $-1.11$ & 5.54 \\

Hexagonal (HX) & $4 $ & $-1.38$ & 5.82 \\

Rectangular (RT-1) & $2$ & $-1.57$ & 6.58 \\

Parallelogram (PR-1) & $2$& $-1.72$ & 6.80\\

Double Rectangular (RT-2) & $3$ & $-2.37$ & 7.23 \\

Double Zigzag (DZZ) & $4$  & $-1.76$ & 7.16 \\

Pentagonal (PNT) & $3$ & $-1.99$ & 7.41 \\

Triple Rectangular (RT-3) & $ 4 $ & $-1.94 $ & 7.02 \\

Centered Hexagonal (CHX) & $5$ & $-2.11$ &  7.57\\ 

Quadruple Rectangular (RT-4) & $5 $ & $-1.88 $ & 7.13 \\

Double Parallelogram (PR-2) & $3$ &  $-1.87$ & 7.68\\ 

Triple Parallelogram (PR-3) & $4$ & $-1.95$ & 7.99\\ 

Quadruple Parallelogram (PR-4) & $5$ & $-1.92$ & 8.12\\ 

\hline
\end{tabular}
\vskip 0.4in
Table-I \\
\vskip 1.3in
\hskip 4.5in V. Kashid \textit{et al.}, Table-I
\end{table}
\end{center}
 
% Table-3 Number of bonds and bonding factor..
\newpage
%\begin{center}
\begin{sidewaystable}[h!b!p!]
\begin{tabular} {l  c c  l  c  c  c } \hline \hline
Geometry & Number & B.E. &\hskip 0.3in Nearest & \hskip 0.2in Bonding Factor & \hspace{0.5cm} Number of \\ 
     & of rows & \hskip 0.1in(eV/atom) &\hskip 0.3in neighbours &\hskip 0.2in ( bonds/atom) & \hspace{0.5cm} cond. channels \\ 
\hline
\hline
Linear (LR) & $1$ &  $-1.70 $ &\hskip 0.4in 2 & 1.00 & 1\\

Zigzag (ZZ) & $2$ &  $ -1.90$ &\hskip 0.4in 2 & 1.00 & 1 \\

Rectangular (RT-1) & $2$ & $-2.09 $ &\hskip 0.4in 3 & 1.50 & 2 \\

Parallelogram (PR-1) & $2$ & $-2.22 $ &\hskip 0.4in  4 & 2.00 &  0\\

Double Rectangular (RT-2) & $3$ &$-2.31 $ &\hskip 0.4in 3, 4 & 1.66 & 3 \\

Pentagonal (PNT) & $3$ & $-2.39$&\hskip 0.4in 3, 4& 1.66 &  2 \\

Double Parallelogram (PR-2) & $3$ &$ -2.46$ &\hskip 0.4in 4, 6 & 2.33 & 2\\ 

Hexagonal (HX) & $4 $ &$ -2.02$ & \hskip 0.4in 2, 3 & 1.25 & 0 \\

Double Zigzag (DZZ) & $4$& $-2.33 $ &\hskip 0.4in 2, 4 & 1.50 & 0 \\

Triple Rectangular (RT-3) & $4$& $-2.40 $&\hskip 0.4in 3, 4 & 1.75 & 4 \\

Triple Parallelogram (PR-3) & $4$ & $-2.40 $ &\hskip 0.4in 4, 6 & 2.50 & 3 \\

Centered Hexagonal (CHX) & $5$ &$ -2.42$ &\hskip 0.4in 3, 5, 6 & 2.20 & 1 \\ 

Quadruple Rectangular (RT-4) & $5$&$-2.43 $ &\hskip 0.4in 3, 4 & 1.80 & 5  \\

Quadruple Parallelogram (PR-4) & $5$ &$-2.43 $ &\hskip 0.4in 4, 6 & 2.80 & 3 \\ 

Bulk gold &  & $-3.16$ &\hskip 0.4in 12 & 6.00 & \\
\hline

\end{tabular}
\vskip 0.05in
Table-II \\
\vskip 0.1in
\hskip 4.5in V. Kashid \textit{et al.}, Table-II
\end{sidewaystable}

%Figures for the paper format are here ------

\newpage
\begin{figure}[h]
{\scalebox{0.73}{\includegraphics{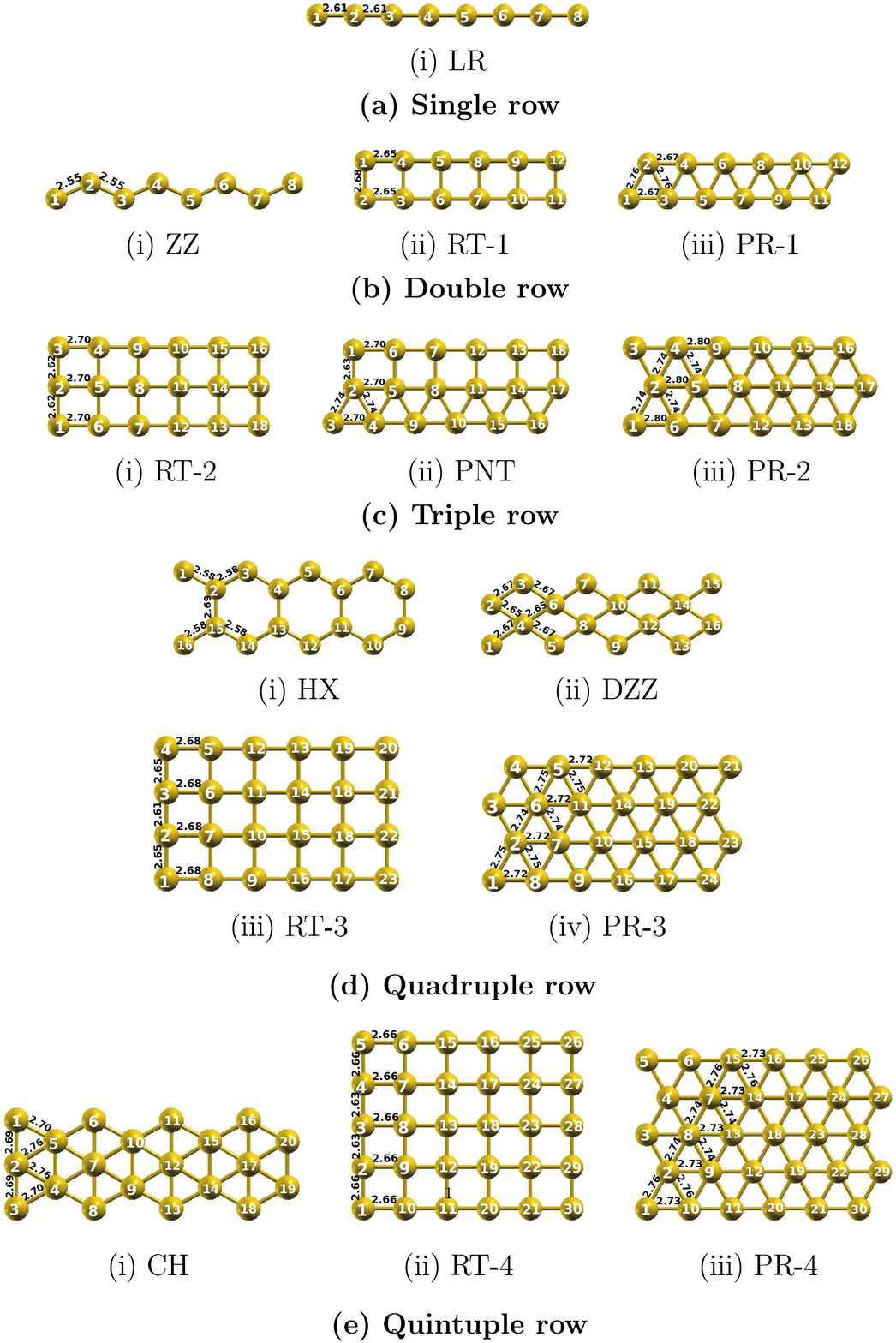}}}\\
\parbox{6in}{\caption{}} \\
\vskip 0.14in
\hskip 3in V. Kashid \textit{et al.}, Fig. 1
\end{figure}

\newpage
\begin{figure}[h]
{\scalebox{0.6}{\includegraphics{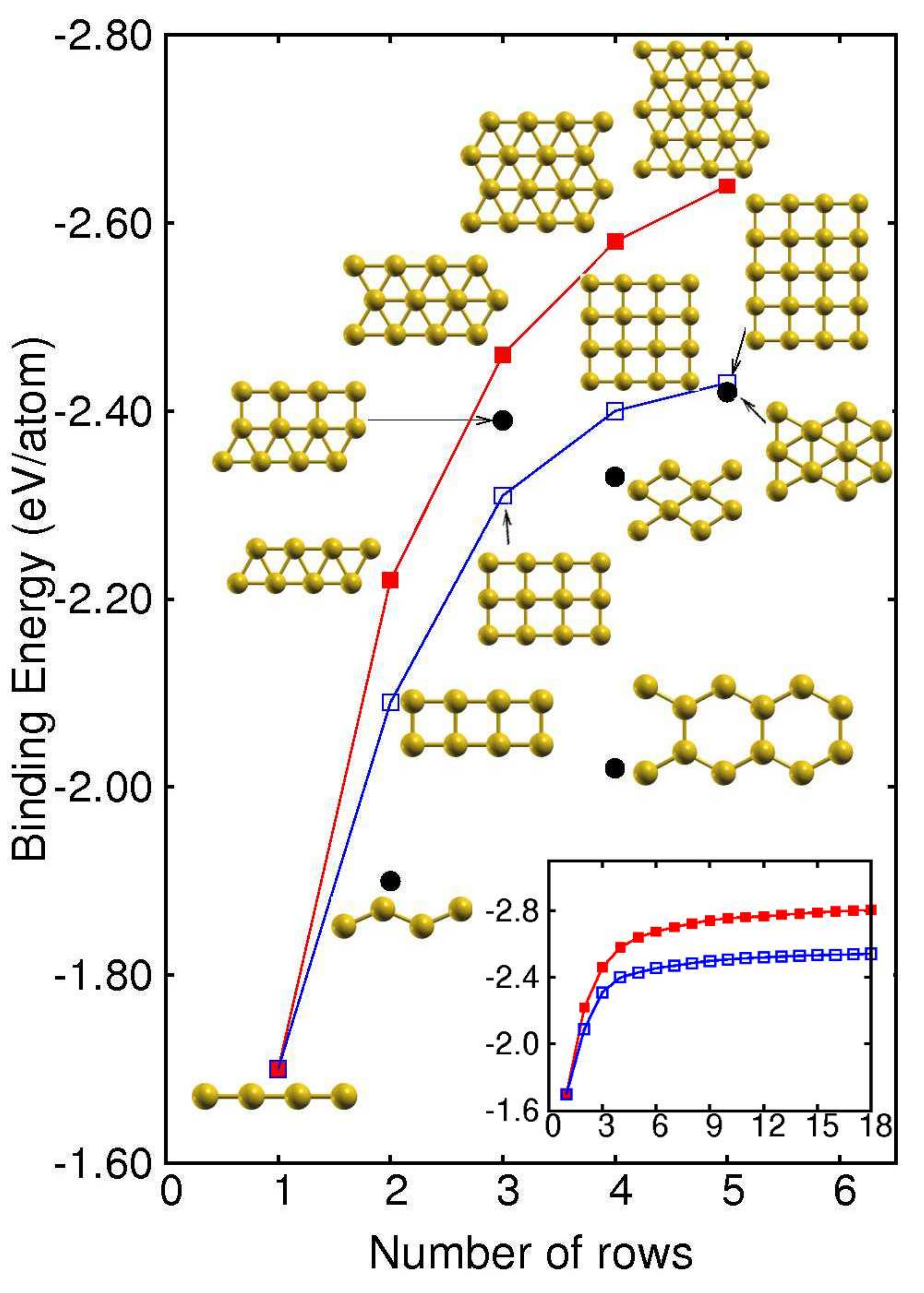}}}\\
\parbox{6in}{\caption{}} \\
\vskip 0.2in
\hskip 3in V. Kashid \textit{et al.}, Fig. 2
\end{figure}

\newpage
\begin{figure}[h]
{\scalebox{0.52}{\includegraphics{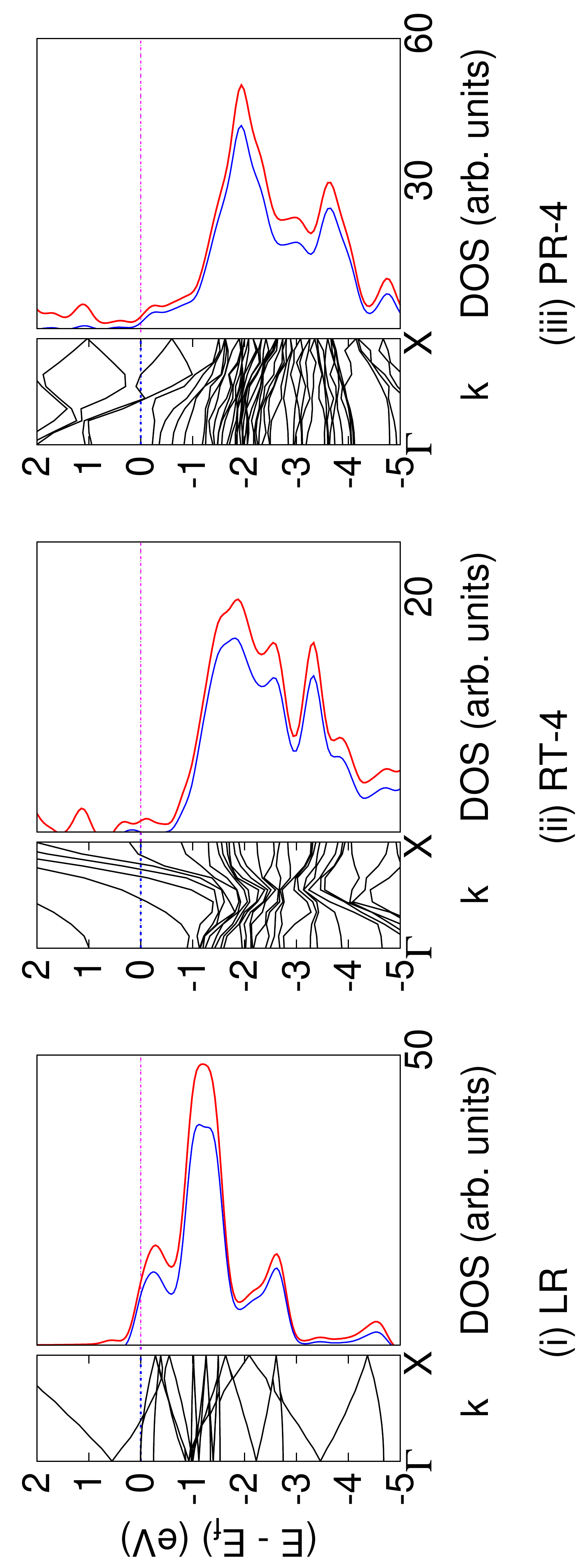}}}\\
\parbox{6in}{\caption{}} \\
\vskip 0.2in
\hskip 3in V. Kashid \textit{et al.}, Fig. 3
\end{figure}

\newpage
\begin{figure}[h]
{\scalebox{0.91}{\includegraphics{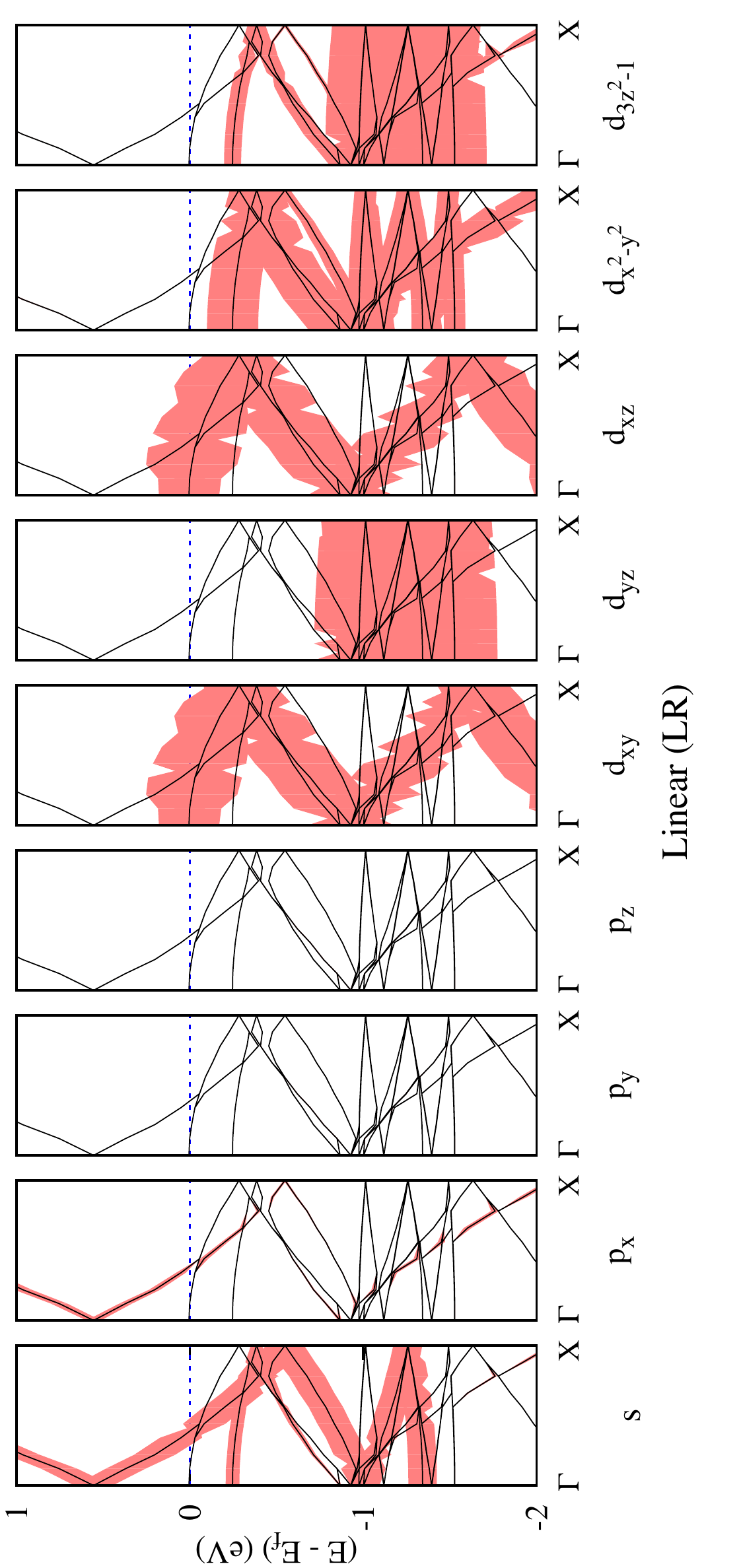}}}\\
%\parbox{6in}{\caption{}} \\
\vskip 0.2in
\hskip 3in V. Kashid \textit{et al.}, Fig. 4(a)
\end{figure}

\newpage
\begin{figure}[h]
{\scalebox{0.91}{\includegraphics{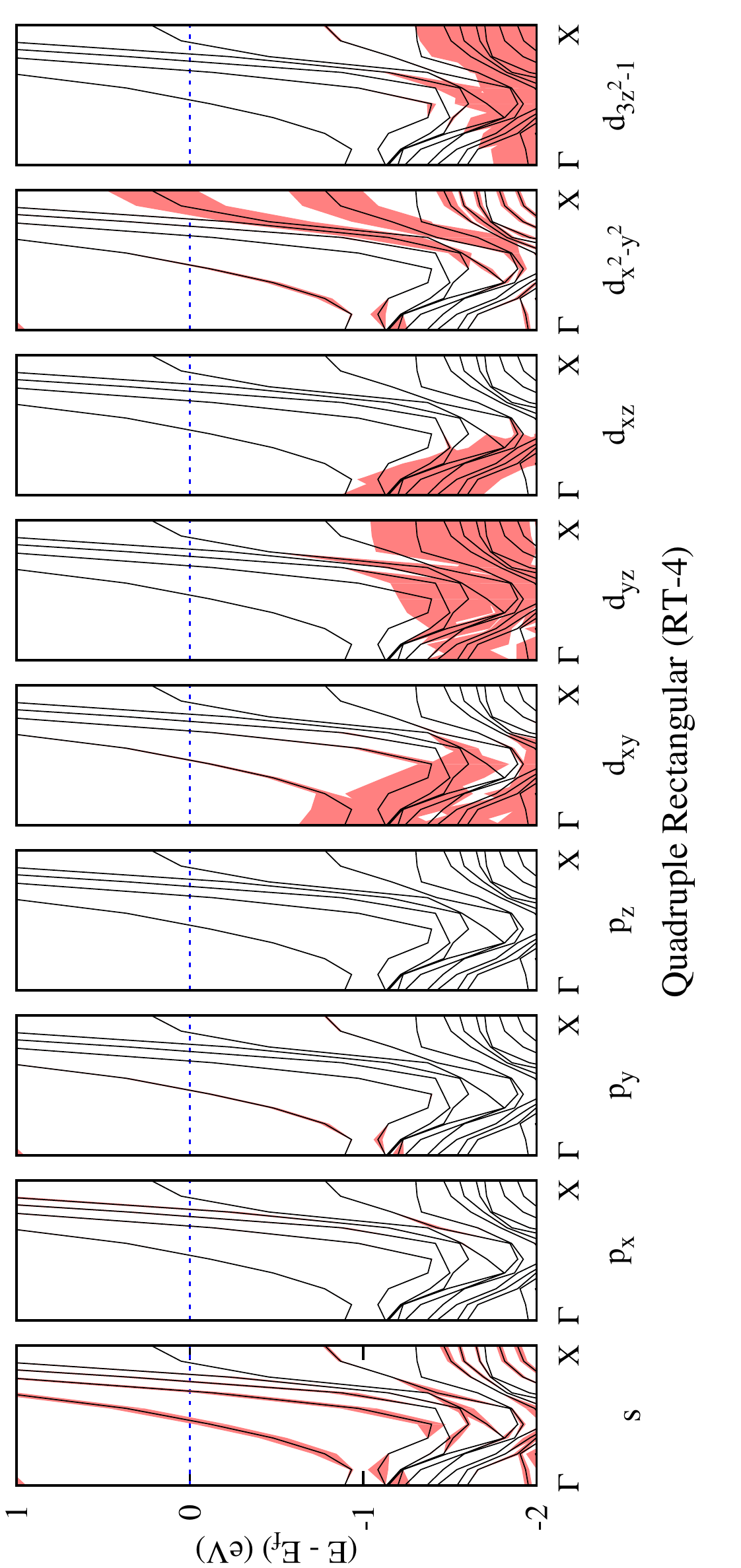}}}\\
%\parbox{6in}{\caption{}} \\
\vskip 0.2in
\hskip 3in V. Kashid \textit{et al.}, Fig. 4(b)
\end{figure}

\newpage
\begin{figure}[h]
{\scalebox{0.91}{\includegraphics{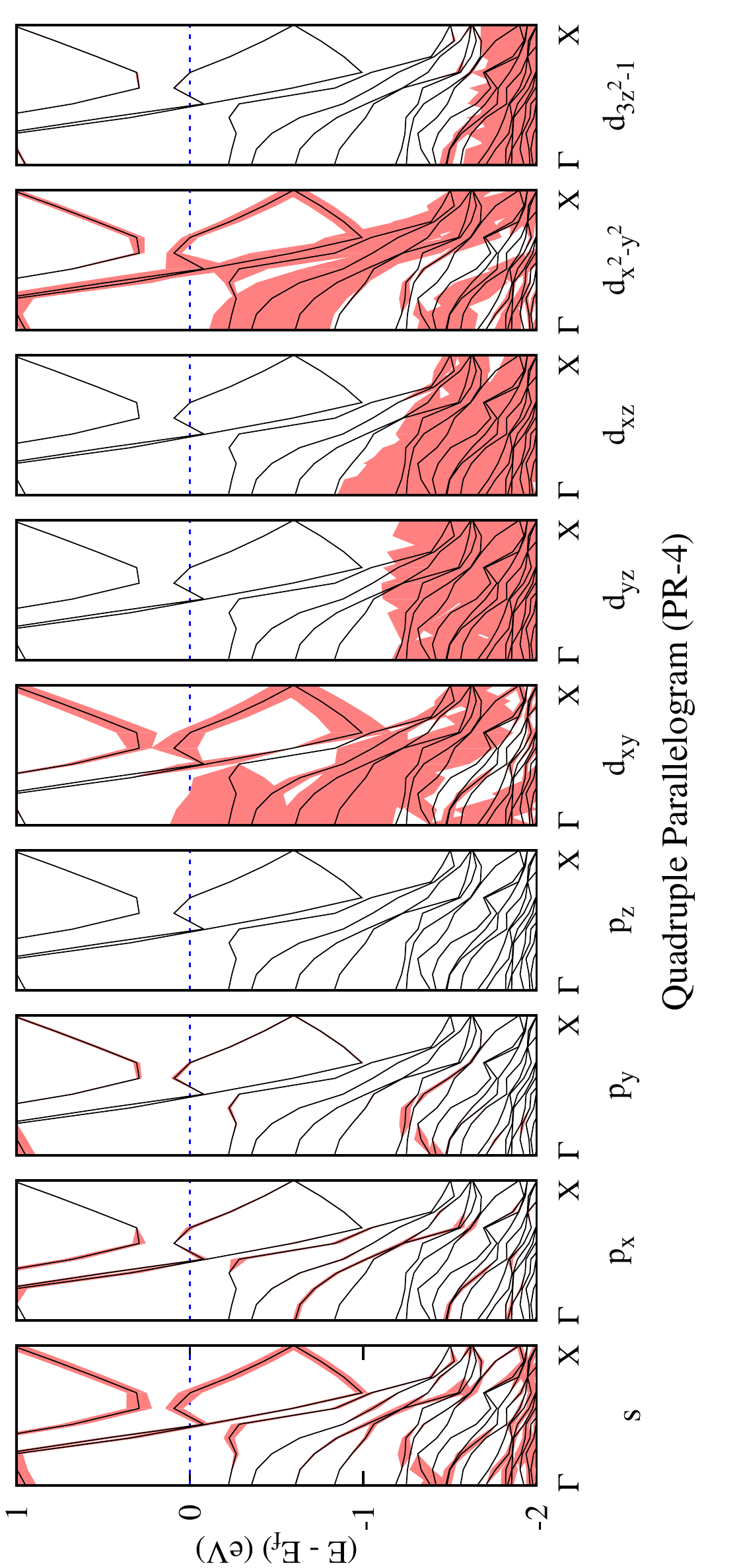}}}\\
\parbox{6in}{\caption{}} \\
\vskip 0.2in
\hskip 3in V. Kashid \textit{et al.}, Fig. 4(c)
\end{figure}

\newpage
\begin{figure}[h]
{\scalebox{0.46}{\includegraphics{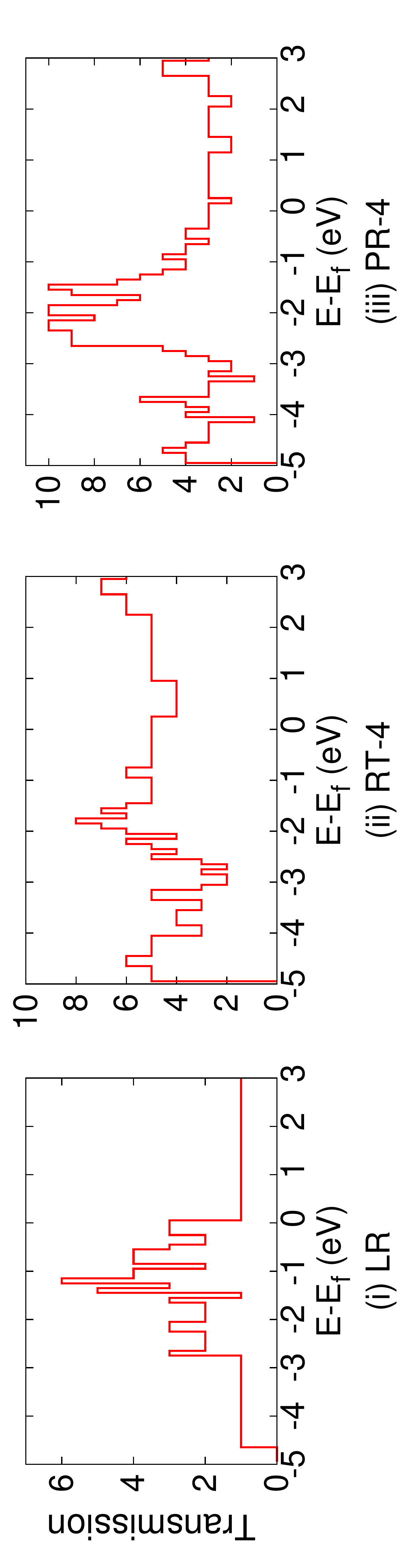}}}\\
\parbox{6in}{\caption{}} \\
\vskip 0.2in
\hskip 3in V. Kashid \textit{et al.}, Fig. 5
\end{figure}

\end{document}